\documentclass[a4paper]{jpconf}
\usepackage{graphicx}
\usepackage{amsmath}
\usepackage{amssymb}

\newcommand{\fig}[1]{Fig.~\ref{fig:#1}}
\newcommand{\htw}{0.47\textwidth}
\newlength{\pictsize}
\begin{document}
\title{Upturn observed in heavy nuclei to iron ratios  by the ATIC-2 experiment}

\author{%
A D Panov$^1$,
N V Sokolskaya$^1$,
V I Zatsepin$^1$,
J H Adams, Jr.$^2$, 
H S Ahn$^3$, 
G L Bashindzhagyan$^1$,
J Chang$^4$,
M Christl$^5$,
A R Fazely$^6$
T G Guzik$^5$,
J Isbert$^5$,
K C Kim$^3$,
E N Kouznetsov$^1$,
M I Panasyuk$^1$,
E S Seo$^3$,
J W Watts$^2$,
J P Wefel$^5$, 
J Wu$^3$
}

\address{$^1$ Skobeltsyn Institute of Nuclear Physics, Moscow State University, Moscow, Russia}
\address{$^2$ Marshall Space Flight Center, Huntsville, AL, USA}
\address{$^3$ University of Maryland, Institute for Physical Science \& Technology, College Park, MD, USA}
\address{$^4$ Purple Mountain Observatory, Chinese Academy of Sciences, China}
\address{$^5$ Louisiana State University, Department of Physics and Astronomy, Baton Rouge, LA, USA}
\address{$^6$ Southern University, Department of Physics, Baton Rouge, LA, USA}

\ead{panov@dec1.sinp.msu.ru}

\begin{abstract}
The ratios of fluxes of heavy nuclei from sulfur ($Z$=16) to chromium ($Z$=24) to the flux of iron were measured by the ATIC-2 experiment. The ratios are decreasing functions of energy from 5 GeV/n to approximately 80 GeV/n, as expected. However, an unexpected sharp upturn in the ratios are observed for energies above 100 GeV/n for all elements from $Z$=16 to $Z$=24. Similar upturn but with lower amplitude was also discovered in the ATIC-2 data for the ratio of fluxes of abundant even nuclei (C, O, Ne, Mg, Si) to the flux of iron. Therefore the spectrum of iron is significantly different from the spectra of other abundant even nuclei. 
\end{abstract}

\section{Introduction}
It is generally expected that the ratio of flux of a secondary cosmic ray nucleus to the flux of the corresponding parent nucleus should be decreasing function of energy. It is a very general consequence of decreasing the diffusion galaxy escape time with energy. This expectation was confirmed very well for energies up to several tens of GeV in HEAO-3-C2 experiment \cite{NUCL-HEAO-1990-AA}, HEAO-3-C3 experiment \cite{NUCL-HEAO-HN-1985-ICRC-28,NUCL-HEAO-HN-1987-ICRC-330,NUCL-HEAO-HN-1988-ApJ} and other experiments. Some essential fractions of the nuclei Ar ($Z$=18) and Ca ($Z$=20) in cosmic rays are supposed to be secondary daughter nuclei of spallation of iron as a parent \cite{SHAPIRO1972-SpRes} therefore Ar/Fe and Ca/Fe ratios may be expected to be decreasing functions of energy. However the data of HEAO-3-C3 experiment \cite{NUCL-HEAO-HN-1985-ICRC-28,NUCL-HEAO-HN-1987-ICRC-330,NUCL-HEAO-HN-1988-ApJ} revealed unexpected upturn of the ratios near the energy 100~GeV/n and their increasing in the region 100--600~GeV/n. The observed upturn was not surely recognized as a real physical effect in \cite{NUCL-HEAO-HN-1985-ICRC-28,NUCL-HEAO-HN-1987-ICRC-330,NUCL-HEAO-HN-1988-ApJ} and possible systematic origin was supposed as well. Similar unexpected upturn was observed in Ti/Fe ratio in the data of the ATIC-2 experiment \cite{ATIC-2009-ZATSEPIN-AstrLett-E} near the energy 70--80~GeV/n, but the statistics for the energies above 100~GeV/n were insufficient and the upturn was considered in \cite{ATIC-2009-ZATSEPIN-AstrLett-E} to be statistically insignificant. While both in the HEAO-3-C3 experiment and in the ATIC-2 experiment the upturn in the ratio was not taken seriously, the resemblance of the observed phenomena forces one to study the phenomenon more accurately. Careful investigation of ratios of heavy nuclei fluxes to the flux of iron in the ATIC-2 experiment is the main object of the present work. Hereinafter we call the range of charges \mbox{$Z$~=~16\,--\,24} as group $H^-$. Manganese ($Z=25$) is not included in the group $H^-$ because it may be contaminated by iron under conditions of the ATIC experiment. Note that there are a number of low-statistics data (emulsion experiments \cite{NUCL-SANRIKU1999-ICRC,EMULS-JACEE-1998-NuclPhysB,EMULS-RUNJOB-2005-ApJ,EMULS-RUNJOB-2005-ICRC-Heavy},  TRACER 2~TeV/n point for B/C ratio \cite{TRACER-2011-ApJ}) that indicate fluxes of secondary nuclei at energies above 100--500~GeV/n to be high relative to simple leaky-box predictions with escape path length $\propto E^{-0.6}$ or to similar models. Therefore the study of secondary to primary fluxes ratio at high energies is an important problem. 

\section{Method}
The details of the ATIC long duration balloon project are descibed in \cite{ATIC-2004-GUZIK-AdvSpRes}. There are two main difficulties to carry out the  described program in the ATIC experiment. The first one is low statistics at high energies (as it was noted in \cite{ATIC-2009-ZATSEPIN-AstrLett-E} in relation to Ti/Fe ratio), the second one is insufficient charge resolution of ATIC experiment in the group $H^-$ that prevents to separate fluxes of adjacent nuclei carefully. The method proposed here solves both problems.

Since the upturn of flux ratios in HEAO-3-C3 (Ar/Fe and Ca/Fe) and in ATIC (Ti/Fe) was observed for quite different charges in the group $H^-$, one reasonably may suppose that the phenomenon is common for the total group $H^-$. Therefore it is meaningful to study the effect averaged on the whole group $H^-$ or for large parts of this group. One improves the statistics by this way and at the same time solves the problem of insufficient charge resolution. If one studies combined spectrum in terms of energy per nucleon for large charge region, then the errors related to the low charge resolution are connected only with the margins of the region and are relatively small.

To proceed with this idea we consider the charge $Z$ to be a continuous variable which is measured by the apparatus. We consider the atomic weight $A$ to be a continuous piece-linear function of $Z$ which is obtained by interpolation of actual relation $Z(A)$. To obtain charges of nuclei we use the usual ATIC's silicon matrix technique with correction for relativistic increasing of ionization capability as described in \cite{ATIC-2009-ZATSEPIN-AstrLett-E}. To reconstruct the primary energy of particles we use the method of differential energy shift described in \cite{ATIC-2007-PANOV-IzvRan}. To reconstruct the spectra at the top of the atmosphere from the spectra at the top of the calorimeter, the correction for fragmentation of nuclei in the atmosphere is carried out with coefficients calculated with FLUKA system for $4.81~g/cm^2$ of the residual atmosphere with technique described in \cite{ATIC-2008-PANOV-ICRC}. These coefficients are 0.897, 0.843, 0.901, 0.840, 0.798, 0.715 for C, O, Ne, Mg, Si and Fe respectively.

\begin{figure}
\begin{minipage}[t]{\htw}
\includegraphics[width=\pictsize]{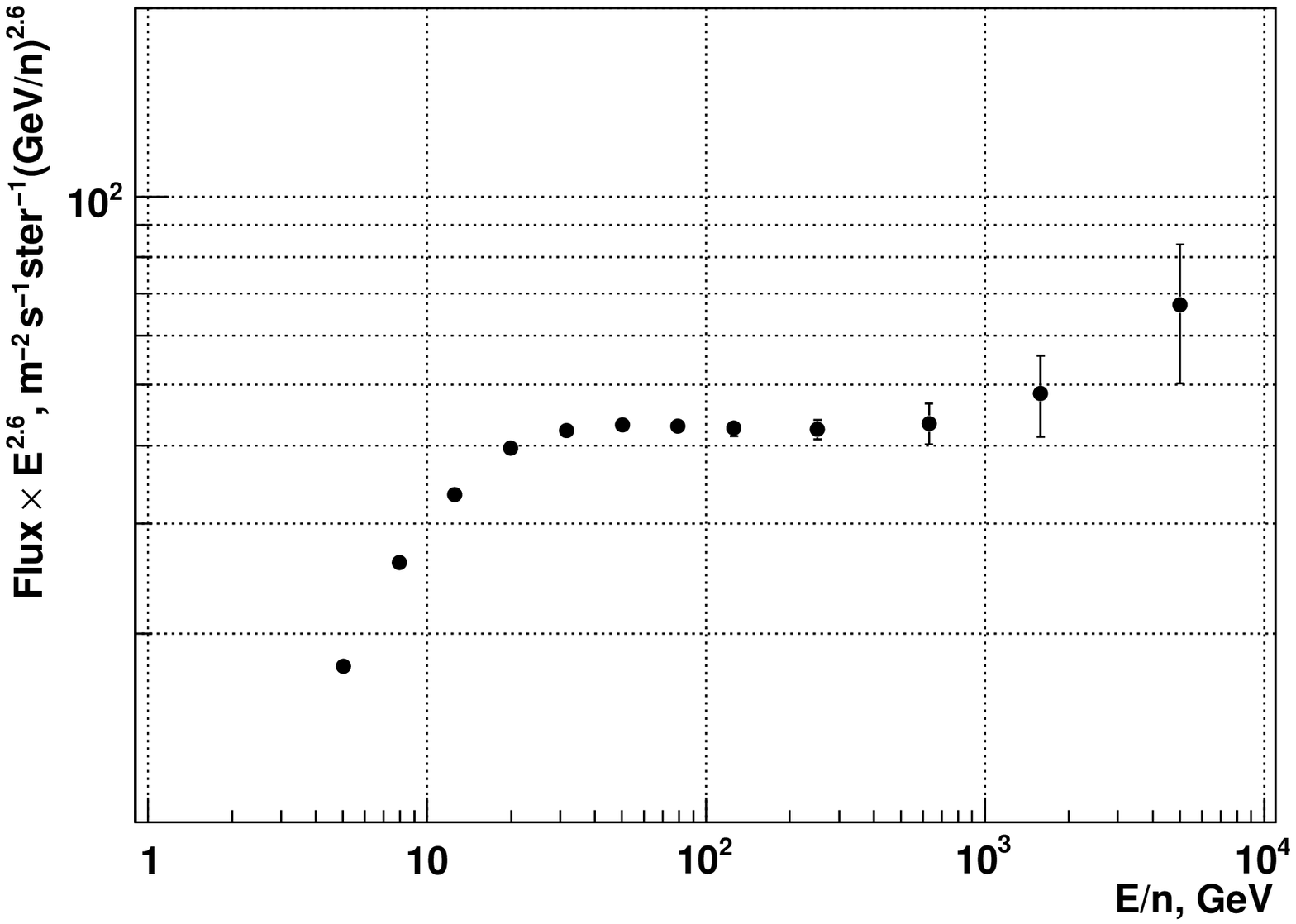}
\caption{\label{fig:AbsAll}The spectrum of all nuclei from carbon to iron $(5.5 \le Z \le 27)$ in the terms of energy per nucleon.}
\end{minipage}\hspace{2pc}%
\begin{minipage}[t]{\htw}
\includegraphics[width=\pictsize]{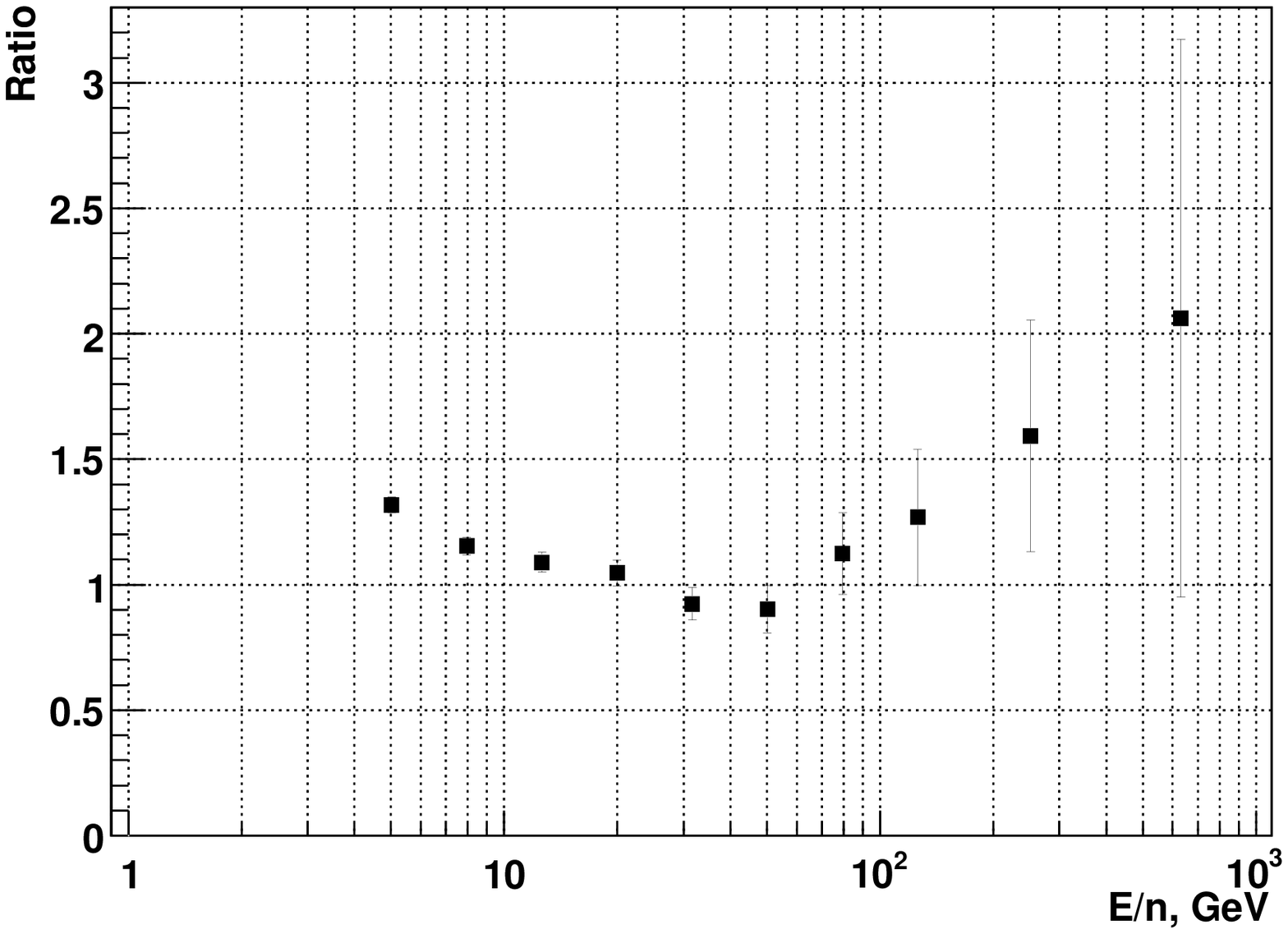}
\caption{\label{fig:Ratio-Whole-Fe} Flux of $(15.5 < Z < 24.5)$  to iron ratio.}
\end{minipage}\\
\begin{minipage}[t]{\htw}
\includegraphics[width=\pictsize]{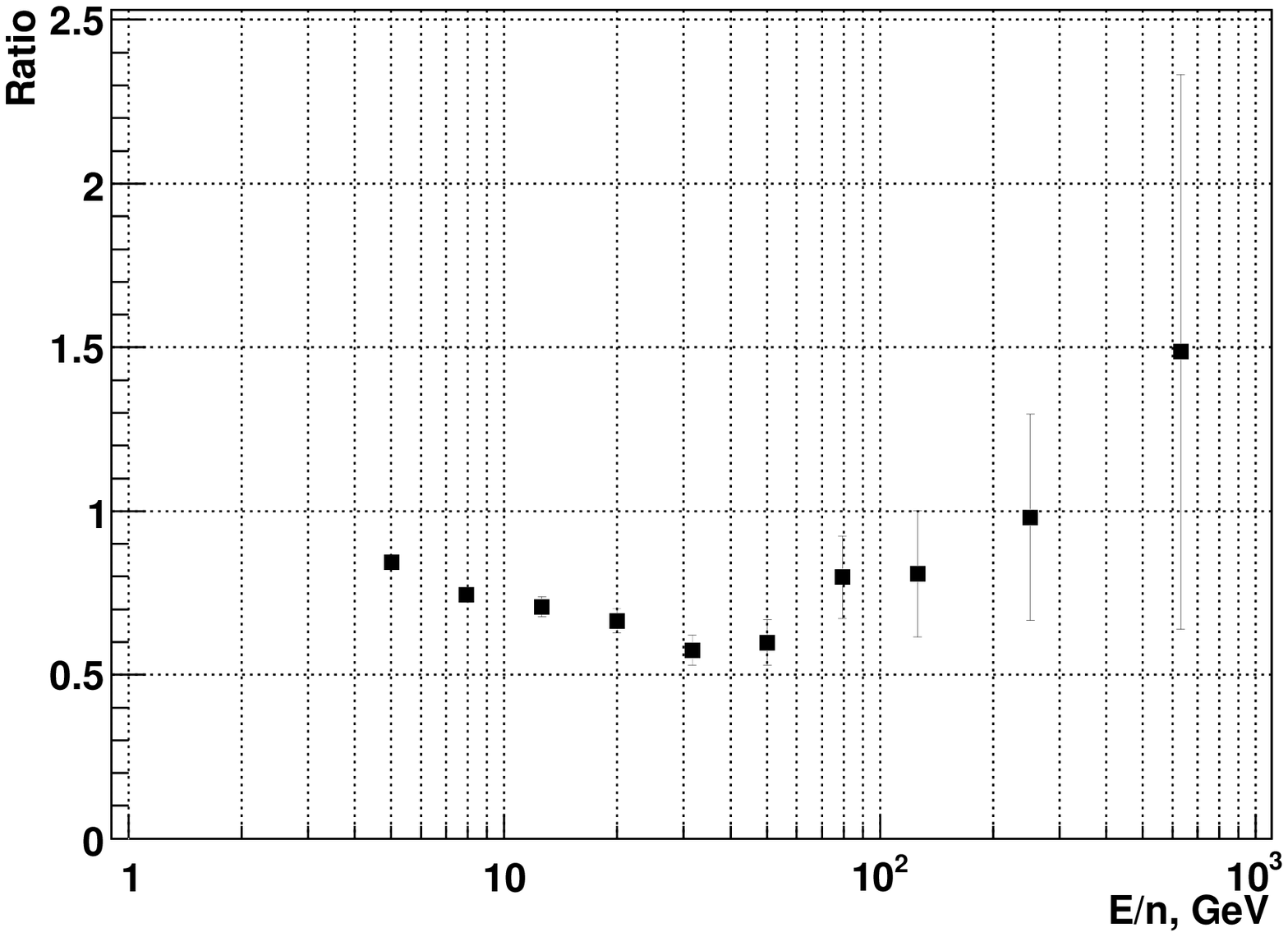}
\caption{\label{fig:Ratio-Whole-Fe-2} Flux of $(15.5<Z<20.5)$ to iron ratio.}
\end{minipage}\hspace{2pc}%
\begin{minipage}[t]{\htw}
\includegraphics[width=\pictsize]{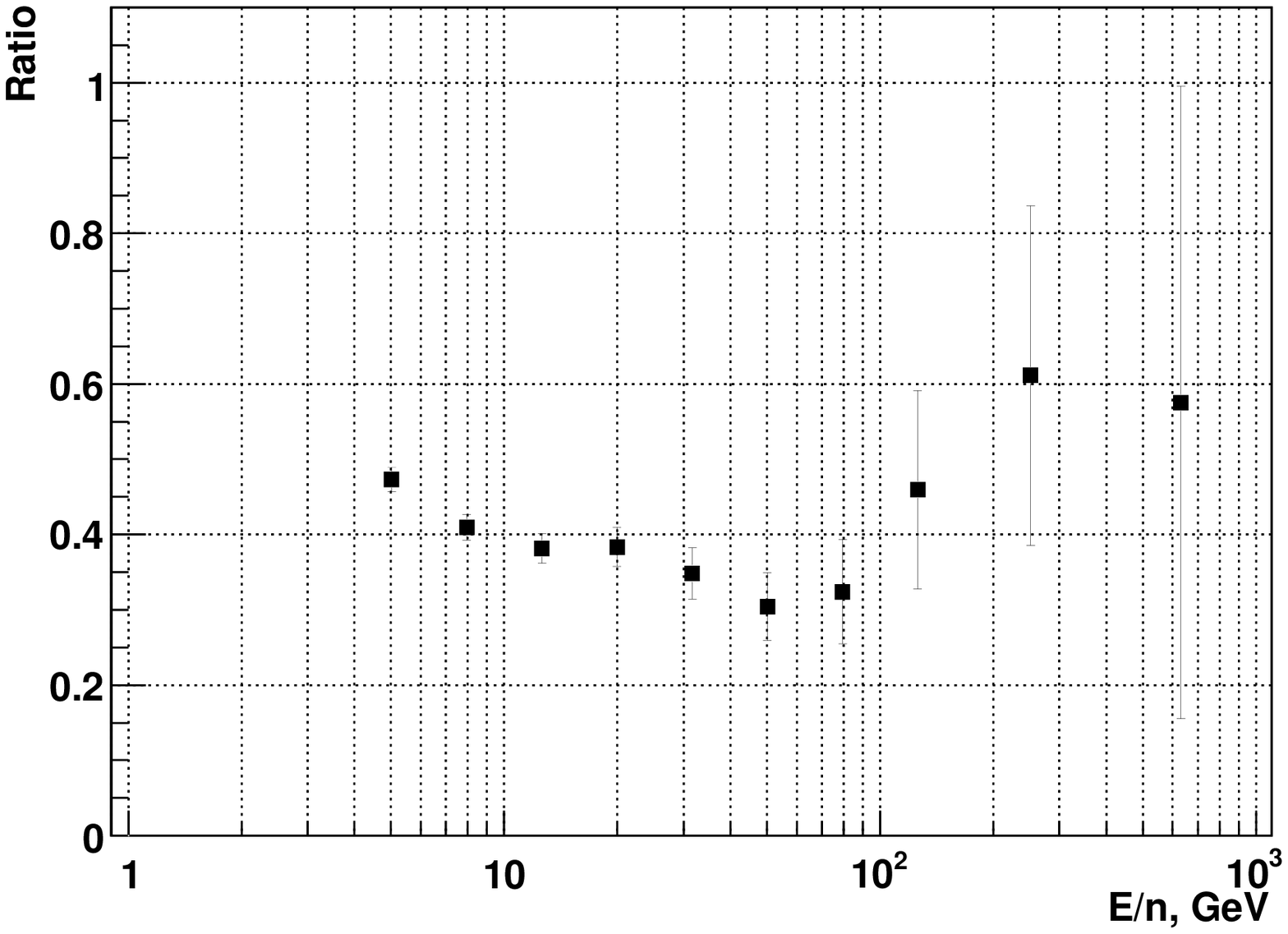}
\caption{\label{fig:Ratio-Whole-Fe-3}Ratio of flux of $(20.5<Z<24.5)$ to iron.}
\end{minipage}
\end{figure}

To check the method we first obtained the summary spectrum for all nuclei from carbon to iron in terms of energy per nucleon. Practically it is the spectrum obtained with the method described above for the charge range $5.5 < Z < 27$. The result is shown in \fig{AbsAll}. The spectrum looks as one could expect. Particularly, the expected hardening of the spectrum at energies above approximately 200~GeV/n discovered by ATIC \cite{ATIC-2007-PANOV-IzvRan} and confirmed by CREAM \cite{CREAM2009-ApJ,CREAM2010A} is clearly seen.

\section{Results}
Ratio of the spectrum for the whole $H^-$ group to the spectrum of iron obtained in the ATIC-2 experiment is shown in \fig{Ratio-Whole-Fe}. Undoubtedly the upturn observed by HEAO-3-C3 for Ar/Fe and Ca/Fe \cite{NUCL-HEAO-HN-1985-ICRC-28,NUCL-HEAO-HN-1987-ICRC-330,NUCL-HEAO-HN-1988-ApJ} is qualitatively confirmed by the ATIC data for the spectrum of $H^-$ group. Ratios of fluxes for sub-regions of $H^-$ group $(15.5 < Z < 20.5)/\mathrm{Fe}$ and $(20.5 < Z < 24.5)/\mathrm{Fe}$ are shown in \fig{Ratio-Whole-Fe-2} and \fig{Ratio-Whole-Fe-3} respectively. The difference between the nature of regions $(15.5 < Z < 20.5)$ and $(20.5 < Z < 24.5)$ is that the contribution of primary nuclei to Ar and Ca fluxes expected to be prominent in the first region, but it is generally supposed that the nuclei from the second region are mainly secondary \cite{SHAPIRO1972-SpRes}. Nevertheless, the upturn is clearly seen for both regions of charges, therefore it looks to be a general property of the whole  group $H^-$.

\section{Discussion}
\begin{figure}
\begin{minipage}[t]{\htw}
\includegraphics[width=\pictsize]{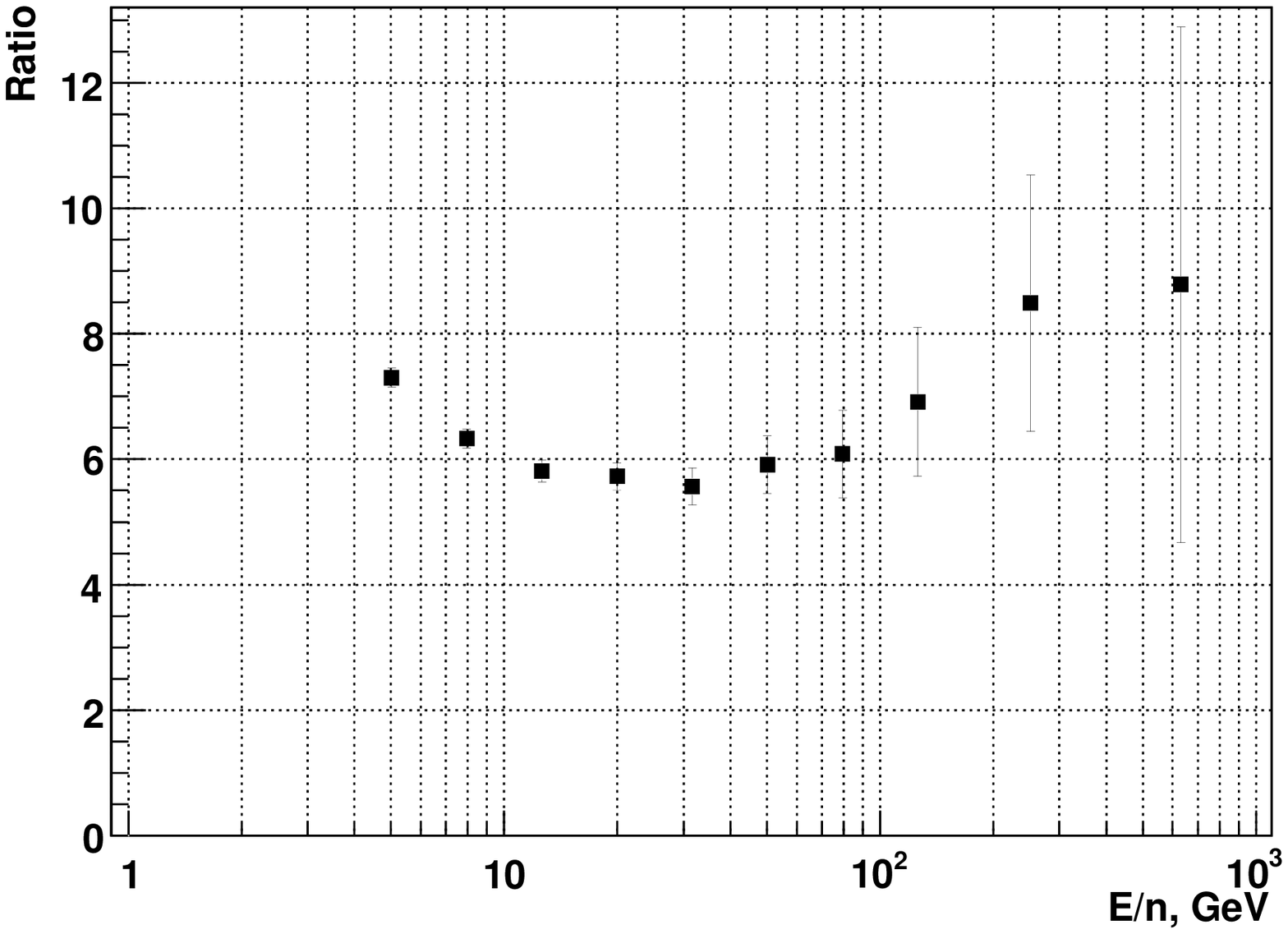}
\caption{\label{fig:O-Fe}Ratio of  oxygen flux to the flux of iron measured by the ATIC-2 experiment.}
\end{minipage}\hspace{2pc}%
\begin{minipage}[t]{\htw}
\includegraphics[width=\pictsize]{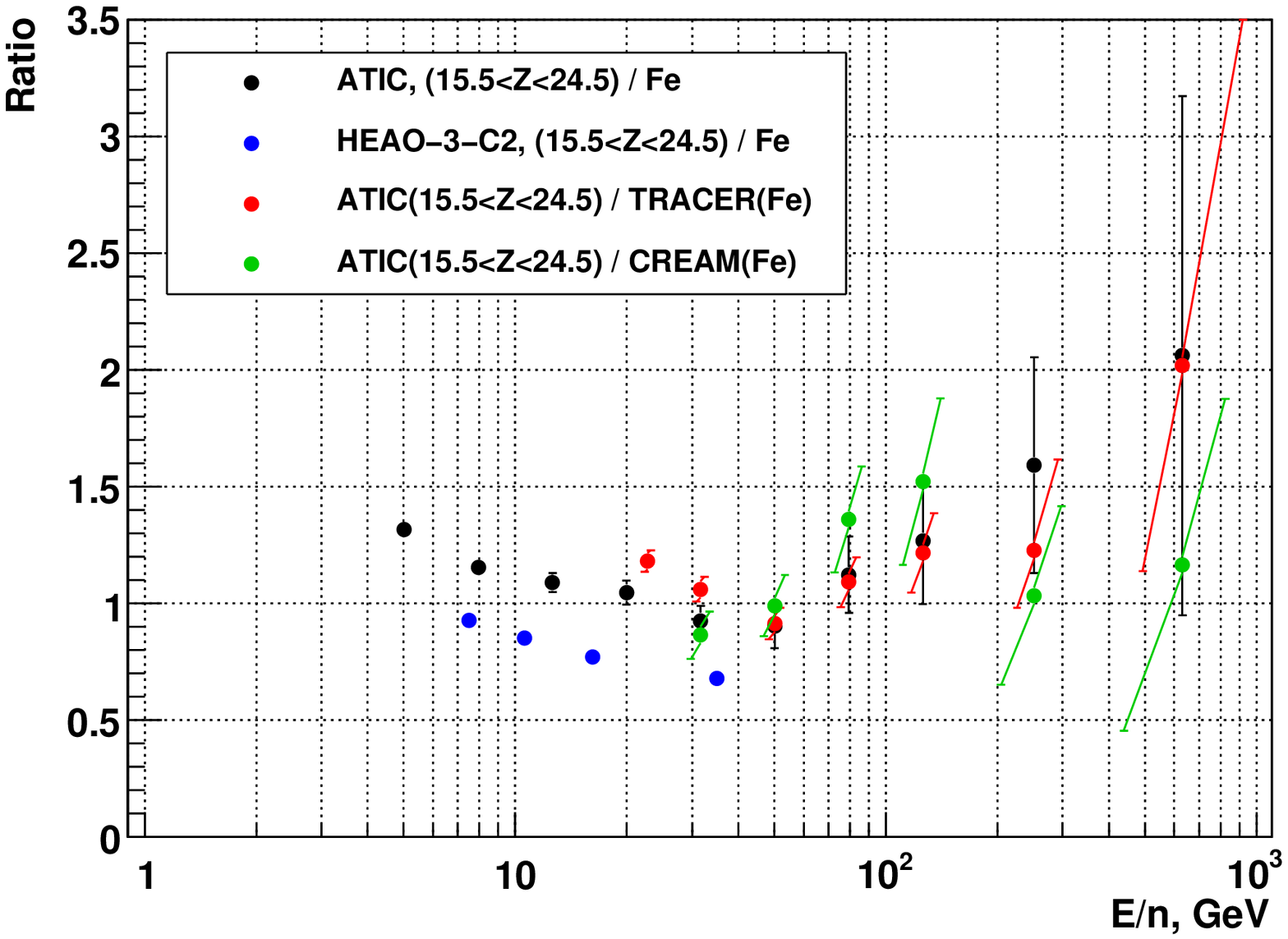}
\caption{\label{fig:Ratio-SubFe-Fe-All} Ratio of $(15.5 < Z < 24.5)$/iron in ATIC and other experiments (see the text).}
\end{minipage}
\end{figure}

From these observations, it would be too early to deduce the conclusion that the upturn of the ratio $H^-$/iron is related to the behavior of spectra of secondary nuclei produced by spallation of iron. The ratio of oxygen to iron fluxes is shown in \fig{O-Fe}. It is seen that there exists similar upturn in the ratio. The ratios of C/Fe, (Ne+Mg+Si)/Fe measured by ATIC-2 look very similar to the O/Fe ratio. Therefore the upturn of ratios is not an attribute of the secondary nuclei exclusively. Rather it may be connected (even if partially) with some special behavior of the iron spectrum in the ATIC data relatively to the spectra of all other nuclei $6 \le Z \le 25$. However, it is important that the amplitude of modulation in the ratios $H^-$/iron (\fig{Ratio-Whole-Fe} -- \fig{Ratio-Whole-Fe-3}) is higher than for the ratios of abundant nuclei like O/Fe (\fig{O-Fe}). Note, that the spectra of abundant even nuclei C, O, Ne, Mg, Si measured by ATIC are in good agreement with the data of other experiments \cite{ATIC-2009-PANOV-IzvRAN-E}.

To exclude the possibility that the observed upturn in the ratio of fluxes of various nuclei to the flux of iron is related to some systematics in the ATIC spectrum of iron, we have calculated the ratio of $H^-$ group measured by ATIC to the iron  measured by TRACER \cite{TRACER-2008B-ApJ} and by CREAM \cite{CREAM2009-ApJ} (there are no direct data for $H^-$/Fe from both TRACER and CREAM). The  TRACER and CREAM data were recalculated to the ATIC energies by interpolation on energy. The result of comparison is shown in \fig{Ratio-SubFe-Fe-All}. The data of HEAO-3-C2 experiment \cite{NUCL-HEAO-1990-AA} are also shown here. All the data are in a reasonable agreement within statistics (an exception is a bit lower absolute ratio of fluxes in HEAO-3-C2, but the shape of the curve reproduces the ATIC's data well). Therefore, there are no signs of systematics in the discovered upturn in the ratio of fluxes. A number of other tests to exclude systematics that we can not mention here were also carried out. Among other factors, we have analyzed the data on super-iron nuclei. No indications of systematics were found. Thus we consider the upturn of ratios to be a real physical phenomenon,  and it seems to indicate that the spectrum of iron is different from other primary nuclei.

The work was supported in Russia by RFBR grant number 11-02-00275.

\section*{References}

\providecommand{\newblock}{}

\end{document}